\documentclass[12pt, preprint]{emulateapj}
\usepackage{natbib}

\def\grtsim{\mathrel{\hbox{\rlap{\hbox{\lower2pt\hbox{$\sim$}}}\raise2pt\hbox{$>$}}}} 
\def\lesssim{\mathrel{\hbox{\rlap{\hbox{\lower2pt\hbox{$\sim$}}}\raise2pt\hbox{$<$}}}}

\def\degree{\nobreak\ifmmode{^\circ}\else{$^\circ$}\fi}

\newcommand{\whz}{W~Hz$^{-1}$}

\newcommand{\lx}{$L_{\rm X}$}

\newcommand{\rtt}{$\langle R \rangle$}

\newcommand{\eff}{$\langle \epsilon \rangle $}

\newcommand{\cbol}{$C_{\nu}$}

\def\lsol{L$_{\odot}$}

\def\msol{M$_{\odot}$}

\newcommand{\densloc}{$\rho_{\bullet ,\rm loc}$}

\shorttitle{Obscured AGNs and the radiation efficiency}
\shortauthors{A. Mart\'\i nez-Sansigre \& A.~M. Taylor }

\begin{document}

\title{The cosmological consequence of an obscured AGN population on the radiation efficiency} 

\author{Alejo Mart\'\i nez-Sansigre}

 \affil{Max-Planck-Institut f\"ur Astronomie, K\"onigstuhl 17,
    D-69117 Heidelberg, Germany; ams@mpia.de}

\author{Andrew M. Taylor}

\affil{Max-Planck-Institut f\"ur Kernphysik, Postfach 103980,
  D-69029 Heidelberg, Germany; andrew.taylor@mpi-hd.mpg.de}

\begin{abstract}  
  In light of recent indications for a large population of obscured
  active galactic nuclei (AGNs), we revisit the mean radiation
  efficiency from accretion onto supermassive black holes (SMBHs),
  \eff, applying a bayesian approach.  We use the integrated comoving
  energy density emitted by AGNs and compare it to the mass density of
  local SMBHs. When considering only optically-selected unobscured
  AGNs, we derive log$_{10}$[\eff]$=$-1.77$^{+0.16}_{-0.11}$ or
  \eff$=$1.7$^{+0.8}_{-0.4}$\%.  Using the AGNs selected using hard
  X-rays, which include unabsorbed and Compton-thin absorbed AGNs, we
  find log$_{10}$[\eff]$=$-1.20$^{+0.15}_{-0.10}$ or
  \eff$=$6.4$^{+2.6}_{-1.3}$\%.  Using optically-selected AGNs, and
  correcting for the obscured population, we inferr
  log$_{10}$[\eff]$=$-1.17$^{+0.11}_{-0.08}$ or
  \eff$=$6.7$^{+1.9}_{-1.1}$\%, which we consider our best
  estimate. We also repeat this calculation for intrinsically luminous
  AGNs ($M_{\rm B}<-23$, quasars), comparing to the SMBH mass density
  in local elliptical galaxies, and find
  log$_{10}$[\eff]$=$-1.27$^{+0.15}_{-0.11}$ or
  \eff$=$5.4$^{+2.2}_{-1.2}$\%.  We discuss how our results can be
  used to estimate the mean spin of accreting SMBHs, both assuming the
  standard thin-disk model of accretion onto black holes and also
  comparing to more recent simulations. Our results can rule out
  maximally rotating SMBHs ($\langle a \rangle=0.998 {G m_{\bullet}
    \over c^{2}}$) at the $\geq$98\% confidence level, as well as high
  rotation values ($\langle a \rangle\geq0.9 {G m_{\bullet} \over
    c^{2}}$) with $\geq$92\% confidence. Our preferred values of
  $\langle a \rangle$ are $\sim$0.25-0.60$ {G m_{\bullet} \over
    c^{2}}$, although even these might be overestimated. Hence, we
  find that on average, SMBHs are not rapidly spinning during
  accretion.  Finally, using an independent measurement of Eddington
  ratios, we estimate the SMBH $e$-folding time for the brightest AGNs
  (quasars) to be $\langle \tau \rangle=100^{+151}_{-60}$~Myr.
 \end{abstract} 
 

\section{Introduction}

Active galactic nuclei (AGNs) are believed to be powered by accretion
onto supermassive black holes (SMBHs), and are expected to leave
behind dormant SMBHs
\citep[][]{1964ApJ...140..796S,1969Natur.223..690L,1984ARA&A..22..471R}.
The observed energy per comoving volume radiated by AGNs, the density
of local SMBHs and the radiation efficiency of accreting SMBHs can all
be related, using conservation of energy, in an elegant argument
proposed by \citet{1982MNRAS.200..115S}.

The original form of the ``So{\l}tan argument'' estimated the density
of relic SMBHs from the distribution of observed quasar fluxes
(magnitudes) and making an educated guess for the
efficiency. Nowadays, SMBHs are believed to reside at the centres of
galaxies, and estimates of the SMBHs masses are possible
\citep[e.g.][]{1995ARA&A..33..581K,1996Natur.383..415E,1998AJ....115.2285M}.
More accurate measurements of the AGN luminosity functions (LFs)
exist, meaning that the least constrained parameter is the radiation
efficiency.  Using variations of the So{\l}tan argument, the
efficiency can be estimated from the comparison between the estimated
total radiated energy from AGNs with the local SMBH
\citep[e.g.][]{2002ApJ...565L..75E,2002MNRAS.335..965Y,2004MNRAS.351..169M,2004MNRAS.354.1020S}.

The discovery of a significant population of obscured AGNs at a range
of redshifts (see Section~\ref{sec:demo}) has important consequences
for the global energetics of quasars and accretion onto SMBHs, given
that the total energy output from the obscured population is
comparable to that of the unobscured population (if not greater).

In this article, we address the issue of how this population can have
a significant effect on the mean radiation efficiency from accretion
onto SMBHs.  We apply a bayesian inference technique to derive
posterior probability distribution functions (PDFs) for the mean
radiation efficiency of SMBHs, \eff.

In Section~\ref{sec:soltan} we outline how the total radiated energy
density and the local SMBHs density can be used to estimate \eff. In
Section~\ref{sec:lf} we discuss the relative contributions of AGNs at
different redshifts to the total radiated energy density.
Section~\ref{sec:demo} summarises the current knowledge on the
obscured AGN population. In Section~\ref{sec:infer} we explain in
detail how we inferr \eff, using a bayesian approach, and how we
combine different estimates of \eff. Section~\ref{sec:spin} discusses
how the mean SMBH spin during periods of accretion can be estimated
from \eff.  Finally, in Section~\ref{sec:efold} we estimate the
$e$-folding time for quasars, and in Section~\ref{sec:conclusion} we
summarise our results.

Throughout this article we refer to powerful AGNs as quasars where,
for unobscured AGNs, the dividing line is generally taken at $M_{\rm
  B}=-23$ ($L_{\nu \rm B}=$$7\times10^{22}$~\whz).  We assume a
$\Lambda$CDM cosmology with the following parameters: $h = H_{0} /
(100 ~ \rm km ~ s^{-1} ~ Mpc^{-1}) = 0.7$; $\Omega_{\rm m} = 0.3$;
$\Omega_{\Lambda} = 0.7$.

\section{The efficiency of accretion onto SMBHs} \label{sec:soltan}

We outline here the main steps of the So{\l}tan argument, in the form that we
will use to inferr the mean radiation efficiency onto SMBHs, \eff.

The total (comoving) energy density radiated by AGNs can be related to
the integrated local SMBH density, \densloc, using three assumptions:
i) that SMBHs grew all their mass by accretion, thus neglecting
mergers between black holes (or rather taking the simplifying
assumption that mergers conserve SMBH mass) ii) that the mean
efficiency \eff\, is constant with luminosity and over the Hubble time
and iii) that AGNs are the manifestation of these periods of growth by
accretion.

For an individual SMBH, accretion with radiation efficiency $\epsilon$
will lead to a bolometric energy:

\noindent \begin{equation}
E_{\rm bol} = \int  \epsilon { {\rm d}m \over {\rm d}t}c^{2}  {\rm d}t ,
\end{equation}

\noindent while a relic SMBH will be left behind, with mass, 

\noindent \begin{equation}
 m_{\bullet , \rm acc}  = \int (1-\epsilon){ {\rm d}m \over {\rm d}t}{\rm
   d}t .
\end{equation}

\noindent The radiated energy no longer contributes to the accreted
mass energy, which the distant observer will see as the SMBH mass,
hence the $(1-\epsilon)$ term.  Rather than tracking individual
sources, we are interested in the relic SMBH mass density
($\rho_{\bullet ,\rm acc}$) inferred from the summed comoving energy
density emitted by AGNs over the age of the Universe, which can be
estimated from the AGN LF.  Converting to observable quantities, we
obtain:

\noindent\begin{equation} \label{eq:integral_no_r}
  \rho_{\bullet , \rm acc}  =  { (1- \langle \epsilon \rangle) \over \langle \epsilon \rangle~c^{2}} \int {{\rm d}t \over {\rm d}z} {\rm d}z \int \it  \,
  C_{\nu} \nu L_{\nu} ~\phi (\it L_{\nu}, \it z){\rm d}L_{\nu} , 
\end{equation} 

\noindent where $\phi (\it L_{\nu}, \it z)$ is the LF for all AGNs,
$C_{\nu}$ is the bolometric correction and \eff\, is the mean
efficiency. The bolometric correction (defined so $L_{\rm bol}=C_{\nu}
\nu L_{\nu}$) is necessary to convert from the monochromatic LF to the
bolometric luminosity density.

If only an unobscured AGN LF is available, one can instead correct for
the missing obscured AGNs:

\noindent\begin{equation} \label{eq:integral}
  \rho_{\bullet , \rm acc}  =  { (1- \langle \epsilon \rangle) \over \langle \epsilon \rangle~c^{2}}(1+\langle R \rangle) \int {{\rm d}t \over {\rm d}z} {\rm d}z \int \it  \,
  C_{\nu} \nu L_{\nu} ~\phi_{\rm u} (\it L_{\nu}, \it z){\rm d}L_{\nu} . 
\end{equation} 

\noindent Here $\phi_{\rm u} (\it L_{\nu}, \it z)$ is the unobscured
AGN LF.  \rtt\, is the mean ratio of obscured to unobscured AGNs so
the term (1+$\langle R \rangle$) is a parametrised way of accounting
for the obscured population.  Estimates exist for the parameters
\cbol\, and \rtt. As mentioned earlier, the least constrained
parameter is \eff, so we treat it here as a free parameter.

\section{Radiated energy density}\label{sec:lf}

A crucial quantity for the estimation of \eff\, is the total energy
density radiated by AGNs, $U$ (the time-integrated luminosity
density).  The AGN population shows a strong evolution so that the
contribution of the low-redshift AGNs ($z\lesssim$0.3) to the energy
density is very modest.
 
Amongst optically-selected AGNs, the luminosity density of AGNs
strongly increases with redshift, with a peak around $z\sim2$-2.5
\citep[e.g.][]{2003A&A...408..499W}, and for the entire AGN population
over cosmic time, $U$ is dominated by the AGNs at $z=$1-3
(particularly those around $L^{\star}$) with the lower-redshift AGNs
contributing a smaller fraction.

\begin{figure} 
\plottwo{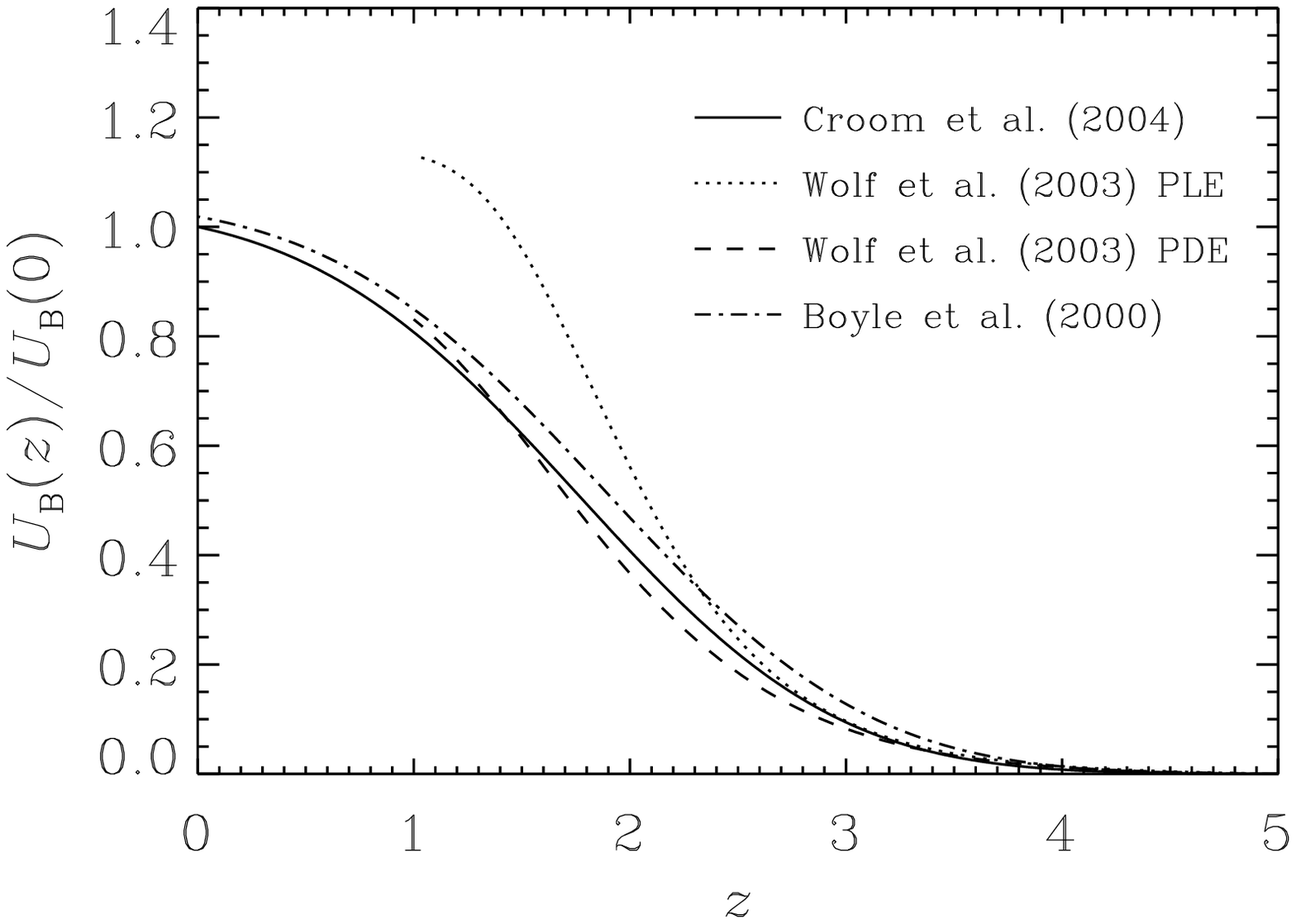}{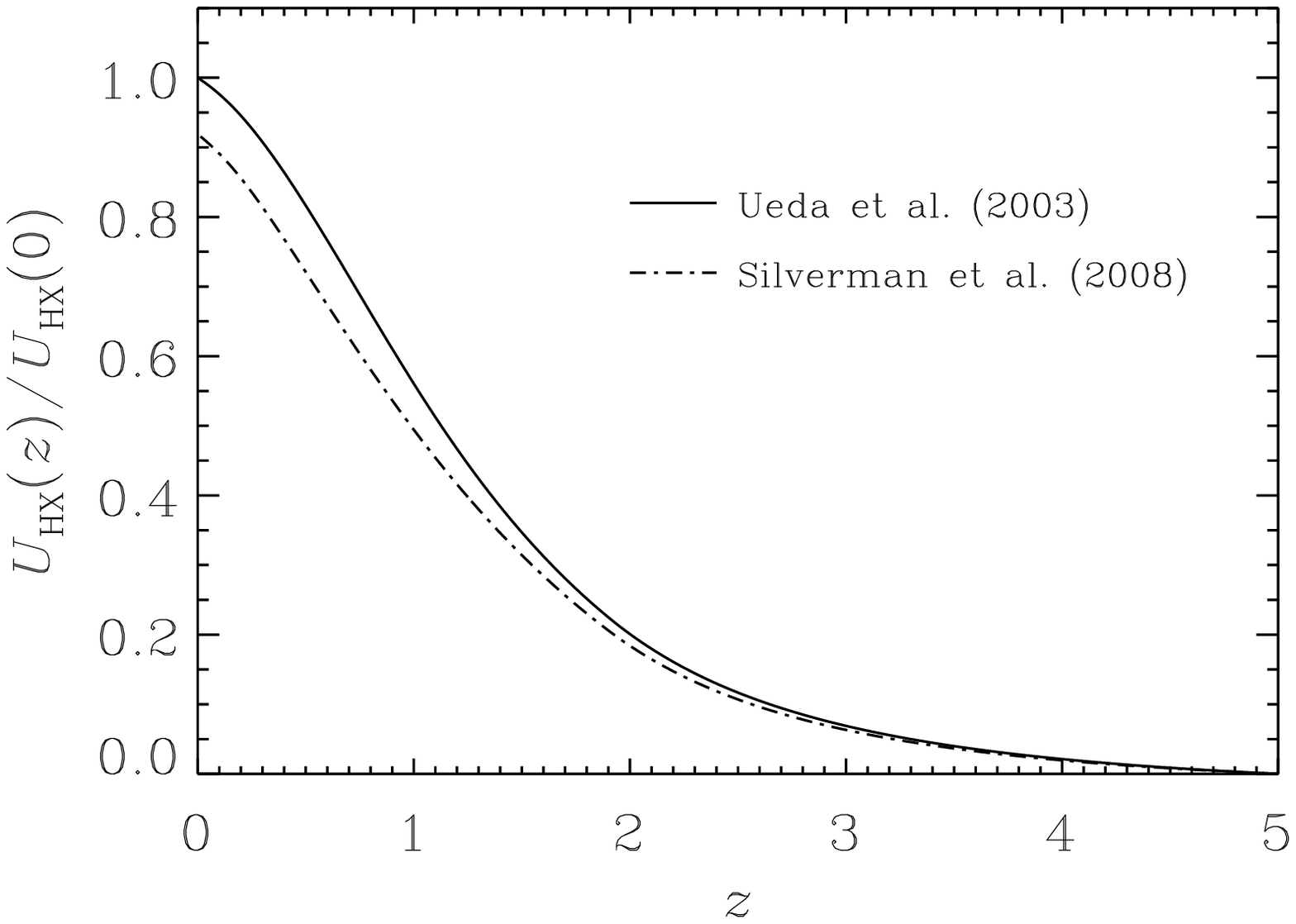}
\caption{\noindent Cumulative energy densities for the optical LFs
  (left) and hard X-ray LF (right). We remove any dependence on
  \cbol\, by dividing them by the maximum value (at $z=0$). The
  maximum values are taken from the \citet{2004MNRAS.349.1397C} LF
  (left) and from the \citet{2003ApJ...598..886U} LF (right). The
  \citet{2003A&A...408..499W} LF, defined only for $z\geq1$, agrees
  well with the other two LFs assuming pure density evolution (PDE). 
  However, pure luminosity evolution (PLE) predicts a somewhat larger
  value for $U_{\rm B}$. The optical LFs include only unobscured AGNs, while
  the X-ray LF also includes Compton-thin absorbed AGNs. }
\label{fig:ut} 
\end{figure}

This is illustrated in the left panel of Figure~\ref{fig:ut}, which
shows the cumulative energy density $U(z)$ [relative to $U(z$$=$0)]
for several optical LFs and one hard X-ray LF.  For the optical
surveys, all defined in the B-band (centred on 4400~\AA, or 2.8~eV),
$U(z)$ is defined as:

\noindent\begin{equation} \label{eq:uto}
  U_{\rm B}(z)  =    \int_{z'=5}^{z'=z} {{\rm d}t \over {\rm d}z'} {\rm d}z' \int \it  \,
   C_{\rm B} \nu_{\rm B} L_{\nu \rm B} ~\phi_{\rm B} (\it L_{\nu \rm B}, \it z'){\rm d}L_{\nu \rm B}.
\end{equation} 

\noindent where $h_{\rm p}\nu_{\rm B}=$2.8~eV ($h_{\rm p}$ is Planck's
constant).  $L_{\nu \rm B}$, $\phi_{\rm B}$ and $C_{\rm B}$ are,
respectively, the luminosity density, LF and bolometric correction at
the B-band.  Plotting $U(z)/U(0)$ removes the dependence on the
bolometric correction and shows the relative contribution at different
epochs.

As can be seen in the left panel of Figure~\ref{fig:ut}, for
optically-selected AGNs, approximately 70\% of $U(z$$=$0) is radiated
between $z=1$ and 3, with 20\% radiated at $z<1$ and 10\% at $z>3$.

However, similar studies undertaken at hard X-ray energies (in the
2-10~keV band) have shown that the lower-luminosity AGNs evolve
differently to high-luminosity AGNs, with their activity peaking at
lower redshifts \citep[e.g.][]{2003ApJ...598..886U}.  Again, we
illustrate the effect by plotting $U(z)$ (Figure~\ref{fig:ut}, right
panel), with $U(z)$ defined as:

\noindent\begin{equation} \label{eq:utx}
  U_{\rm X}(z)  =    \int_{z'=5}^{z'=z} {{\rm d}t \over {\rm d}z'} {\rm d}z' \int \it  \,
   C_{\rm X}  L_{\rm X} ~\phi_{\rm X} (\it L_{\rm X}, \it z'){\rm d}L_{\rm X}.
\end{equation} 

\noindent where $L_{\rm X}$, $\phi_{\rm X}$ and $C_{\rm X}$ are the
hard X-ray luminosity, LF and bolometric correction, but note that
here $L_{\rm bol}=C_{\rm X} L_{\rm X}$.

From the right panel of Figure~\ref{fig:ut} we can see that only about
50\% of $U$ was radiated between $z=1$ and 3, with $\sim$45\% radiated
at $z<1$ and $\sim$5\% at $z>3$.

Thus, from optically-selected LFs one expects the AGNs at $1 \leq z
\leq 3$ to dominate, although the hard X-ray LF indicates that the
contribution of $z < 1$ AGNs is similarly important and should not be
overlooked.

\section{Demography of obscured AGNs}\label{sec:demo}

Obscured AGNs are believed to be accreting SMBHs where, between X-ray
and near-infrared wavelengths, gas and dust block the line of sight to
the central region. The strong continuum from the accretion disk and
the broad emission lines suffer from heavy extinction by dust, while
the narrow emission lines may sometimes be visible, depending on the
obscuring geometry.  Photoelectric absorption by the intervening gas
and dust will supress the soft X-ray emission ($\lesssim$5~keV). If
the column density is large enough to be optically-thick to Compton
scattering\footnote{A source is Compton-thick if the absorbing column
  density, $N_{\rm H}$, is $\geq {1/\sigma_{T}}=$$1.5\times10^{28}$
  m$^{-2}$, where $\sigma_{T}$ is the Thomson electron scattering
  cross-section. X-ray absorbed, Compton-thin sources have $10^{26}$
  m$^{-2} \lesssim N_{\rm H} \lesssim 10^{28}$ m$^{-2}$, and X-ray
  unabsorbed sources have $N_{\rm H} \lesssim 10^{26}$ m$^{-2}$}, then
even the hard X-rays ($\grtsim$5~keV) will be heavily affected.

The obscured AGNs are indistinguishable from galaxies in optical and
near-infrared imaging surveys, which has made their identification
significantly harder than that of their unobscured counterparts. There
are mainly four methods for looking for these objects, which we
summarise in chronological order, and we concentrate here on the most
recent results.

Radio selection (wavelength $\grtsim$6~cm, energy
$\lesssim$20~$\mu$eV). Dust and gas are transparent to moderately
high-frequency radio waves, so radio-loud obscured AGNs can be readily
identified, in the form of narrow-line radio galaxies.  Amongst the
radio-loud population, the obscured to unobscured ratio is found to be
$\sim$1:1 \citep{2000MNRAS.316..449W}. However, the fraction of
unobscured AGNs that are radio loud is small \citep[$\sim$10\%,
e.g. ][]{2002AJ....124.2364I}, so that the narrow-line radio galaxies
are expected to represent a similarly small fraction of the obscured
AGN population \citep[e.g.][]{2004AJ....128.1002Z}.
  
Selection at hard X-ray energies ($\sim$2-10~keV). This method is
sensitive to the Compton-thin absorbed AGN population, and it has
identified large numbers of sources in deep X-ray surveys
\citep[see][for a review]{2005ARA&A..43..827B}. \citet
{2007A&A...463...79G} find the ratio of Compton-thin absorbed to
unabsorbed AGNs to vary from $\sim$1:1 at the high luminosity end
(\lx$\sim10^{12}$~\lsol) to $\sim$4:1 at the low luminosity end
(\lx$\sim10^{8}$~\lsol). Using the hard X-ray background (at energies
$\sim$3-100~keV) as an extra constraint, these authors find that the
Compton-thick and Compton-thin absorbed populations are likely to have
similar numbers (i.e $\sim$2:1 at high luminosities and up to
$\sim$8:1 at low luminosities).

Selection using optical spectroscopy (wavelengths $\sim$3500-7500~\AA,
energies $\sim$1.5-3.5~eV). If the geometry of the dust is such that
the narrow emission lines are not obscured, these are detectable via
optical spectroscopy. Using the spectroscopic database from the Sloan
Digital Sky Survey, \citet {2003AJ....126.2125Z} were able to select
large numbers of obscured luminous AGNs (obscured quasars). Using the
[O~III] 5007~\AA\, line strength as a isotropic proxy for the
intrinsic AGN luminosity, the ratio of spectroscopically-selected
obscured quasars to unobscured quasars is $\sim$1.2-1.5:1 \citep[][
for $L_{\rm [O~III]}\geq10^{9}$~\lsol]{2008arXiv0801.1115R}. At lower
[O~III] luminosities ($L_{\rm [O~III]}\sim 10^{7}$~\lsol) \citet
{2005MNRAS.360..565S} estimate the obscured to unobscured ratio to be
$\sim$4:1.

Mid-infrared selection (wavelengths $\sim$3-30~$\mu$m, energies
0.04-0.4~eV).  The extinction due to dust becomes very small at these
wavelengths, so mid-infrared selection should be sensitive to the
radio-quiet population, including obscured AGNs showing no narrow
lines.  Recent work has suggested it is also sensitive to the
Compton-thick population
\citep[e.g.][]{2006ApJ...642..673P,2007MNRAS.379L...6M}. However, care
must be taken to separate starburst contaminants from real AGNs, since
these star-forming galaxies also show bright mid-infrared emission.
Combining mid-infrared and radio criteria, \citet{2008ApJ...674..676M}
found a $\sim$2:1 ratio at the high-luminosity end ($\nu L_{\nu
  \rm~8~\mu m}\grtsim 10^{12}$~\lsol), while
\citet{2005ApJ...634..169D} found a $\sim$4:1 ratio amongst lower
luminosity AGNs ($\nu L_{\nu \rm~24~\mu m}\sim 10^{10}$~\lsol).

Based on mid-infrared selection with no radio criteria,
\citet{2007AJ....133..186L} found a $\sim$2:1 ratio amongst bright
AGNs (with a median value $\nu L_{\nu \rm~5~\mu m}\sim
5\times10^{11}$~\lsol). Using a power-law criterion, \citet
{2006ApJ...640..167A} also found a $\sim$2:1 ratio amongst powerful
AGNs (70\% of their sources having $L_{\rm IR}\geq 10^{12}$~\lsol)
while \citet{2007ApJ...660..167D} find a 4:1 ratio for less luminous
sources (with typically $L_{\rm X}\lesssim 5\times 10^{9}$~\lsol\, or
expected total infrared luminosities $L_{\rm IR}\lesssim
10^{11}$~\lsol) but only 2:1 if the sources are required to be X-ray
detected.

A possible picture fitting these observations is the following:
amongst luminous AGNs (quasars), the obscured to unobscured ratio
seems to be $\sim$2:1, with about half of the obscured quasars being
Compton-thick \citep{2007A&A...463...79G,2007MNRAS.379L...6M}, and
about half showing no narrow emission lines \citep[][which explains
the $\sim$1:1 ratio found from spectroscopically-selected samples by
Reyes et al., 2008]{2008ApJ...674..676M}. Amongst radio-loud quasars,
the obscured to unobscured ratio is $\sim$1:1 with almost all obscured
quasars (narrow-line radio galaxies) showing narrow emission lines.
An explanation for this is that the lack of narrow lines is due to
kpc-scale dust \citep[e.g.][]{2006MNRAS.370.1479M,2006ApJ...645..115R}
and radio jets are efficient at clearing dust
\citep{2002ApJ...568..592B,2002MNRAS.331..435W}, so the fraction of
radio galaxies obscured by kpc-scale dust is small.

Amongst the low-luminosity AGNs, from X-ray, optical spectroscopy and
mid-infrared studies the obscured to unobscured ratio seems to be
$\sim$4:1, consistent with the ratio inferred in the local Universe
\citep[e.g.][]{1999ApJ...522..157R}.  However, modelling of the hard
X-ray background allows this ratio to be as high as 8:1 provided that
half of the absorbed AGNs are Compton-thick
\citep{2007A&A...463...79G}.

All the evidence points towards an obscured fraction that increases
with decreasing AGN luminosity \citep[which was noted a long time ago,
see ][]{1982ApJ...256..410L}. Whether the obscured fraction evolves or
not with redshift is still unclear \citep[see
e.g.][]{2003ApJ...598..886U,2006ApJ...652L..79T,2007A&A...463...79G}. The
studies cited above range from $z\leq0.3$ to $z\sim2$, and luminous
AGNs are present in the entire range
\citep{2003AJ....126.2125Z,2007AJ....133..186L,2008ApJ...674..676M}.
Lower-luminosity AGNs, however, are typically only found at low and
intermediate redshifts ($z\lesssim$1), as expected from flux-limited
samples.  Thus, it is difficult at this stage to judge whether the
obscured fraction varies with redshift.

\section{Inferring  \eff}\label{sec:infer}

Having discussed the radiated energy density from AGNs and the
inferred ratio of obscured to unobscured sources, we proceed to use
these to estimate the mean radiation efficiency of AGNs. We will
inferr different values of \eff, from the optical LF, with and without
adding obscured AGNs, as well as the hard X-ray LF. This will
illustrate the effect on \eff\, of the obscured AGN population.

Under the assumption that the local SMBHs
grew by accretion:

\begin{equation}
\rho_{\bullet ,\rm loc}=\rho_{\bullet ,\rm acc}
\end{equation}

\noindent With $\rho_{\bullet ,\rm acc}$ given by
Equations~\ref{eq:integral_no_r} or \ref{eq:integral}, we can estimate
the value of \eff\, which yields the closest agreement between
$\rho_{\bullet ,\rm loc}$ and $\rho_{\bullet ,\rm acc}$.  For
convenience we define $u\equiv U/C_{\nu}$, where $U$ is defined by
Equation~\ref{eq:uto} or \ref{eq:utx}.  This is done to keep separate
and explicit the uncertainties in the LF (represented by $u$) and on
the bolometric correction (\cbol).

We now apply Bayes' theorem, to obtain the posterior PDF of \eff,
given the measurement of \densloc\, in the local Universe and
marginalising over the parameters $u$, \rtt\, and \cbol:

\begin{eqnarray}\label{eq:post}
{\rm P(\langle \epsilon \rangle|~\rho_{\bullet,\rm loc})} = 
 {{\rm P(\langle \epsilon \rangle)} \over {\rm P(\rho_{\bullet,\rm
       loc})}}~{\rm P(\rho_{\bullet,\rm loc}|~\langle \epsilon \rangle)} =
 \nonumber \\
 {{\rm P(\langle \epsilon \rangle)}\over {\rm P(\rho_{\bullet,\rm
       loc})}} \int   {\rm P({\it u})}  {\rm
   d}u \times \nonumber\\\int   {\rm P({\it \langle R \rangle})}  {\rm
   d}\langle R \rangle  \int   {\rm
   P({\it C_{\nu}})} {\rm P(\rho_{\bullet,\rm loc}|~\langle \epsilon \rangle,
   {\it u},  {\it \langle R \rangle},  {\it C_{\nu}})}{\rm d}C_{\nu}.\,\,\,\,\,\,\, 
\end{eqnarray}

\noindent For our purpose, \densloc\, is our data, so that P(\densloc)
is the evidence, P($u$), P(\rtt) and P(\cbol) are the prior PDFs for
$u$, \rtt\, and \cbol\, respectively. We have made use of the fact
that our parameters are independent, so that P(\rtt $|$
\eff)$=$P(\rtt), P($u$$|$ \eff)$=$P($u$) and P(\cbol $|$
\eff)$=$P(\cbol). P(\densloc$|$ \eff, $u$, \rtt, \cbol) is the
likelihood of \densloc\, given \eff, \rtt\, and \cbol:

\begin{eqnarray}\label{eq:like} 
{\rm P(\rho_{\bullet,\rm loc}|~\langle \epsilon \rangle, {\it u},  {\it \langle R
    \rangle},  {\it C_{\nu}})} = 
{1 \over (2\pi)^{1/2} \sigma_{\rho_{\bullet ,\rm loc}} }e^{-{1\over2}\left(
      { \rho_{\bullet ,\rm loc}-\rho_{\bullet ,\rm acc} \over \sigma_{\rho_{\bullet ,\rm loc}}}\right)^{2}}.
\end{eqnarray}

\noindent  From
Equation~\ref{eq:integral} and the definition of $u$,  $\rho_{\bullet ,\rm acc}$ is:

\begin{equation}\label{eq:densacc_no_r}
\rho_{\bullet ,\rm acc} = { (1- \langle \epsilon \rangle) \over \langle
  \epsilon \rangle~c^{2}} C_{\nu} u,
\end{equation}

\noindent or: 

\begin{equation}\label{eq:densacc}
\rho_{\bullet ,\rm acc} = { (1- \langle \epsilon \rangle) \over \langle
  \epsilon \rangle~c^{2}}(1+\langle R \rangle)C_{\nu} u,
\end{equation}

\noindent depending on whether obscured AGNs are to be added on via
the (1$+$\rtt) term. Note that we have made the simplifying assumption
that \rtt\, is independent of luminosity, this will be addressed in
Section~\ref{sec:best}.

\subsection{Priors}\label{sec:priors}

We now assign a prior PDF to each of the parameters \cbol, \rtt, $u$,
and \eff, considering three scenarios: a) considering only unobscured
AGNs b) considering the X-ray absorbed and unabsorbed AGNs, but not
the Compton-thick ones and c) attempting to include all obscured and
unobscured AGNs d) same as c) but considering only the bright AGNs
(quasars).

\subsubsection{{\rm P(\cbol)}}

In this work we are interested in the bolometric output from accretion
onto the SMBH, which we assume produces X-ray, optical and ultraviolet
light predominantly and emits isotropically.  We also assume that the
dust responsible for the observed infrared emission (and the
orientation-dependent SED) is on a larger scale, and not directly
relevant to the accretion disk.  Thus, the bolometric corrections for
accretion onto the SMBH must convert from B-band luminosity to
bolometric luminosity integrated over the optical, ultraviolet and
X-ray range only.

We adopt the luminosity-dependent bolometric corrections of
\citet{2004MNRAS.351..169M} and \citet{2007ApJ...654..731H}, both of
which follow these assumptions.  We assume gaussian priors around
these values, with the former having typical uncertainties
${\sigma_{\rm C_{\rm bol}} \over C_{\rm bol}}=$0.05 and 0.10 for the
B-band and hard X-rays, respectively, and the latter ${\sigma_{\rm
    C_{\rm bol}} \over C_{\rm bol}}=$0.08 and 0.11 for the same bands.

\subsubsection{{\rm P($u$)}}\label{sec:pu}

We estimate the total radiated energy density from unobscured AGNs by
using the B-band LF, and the total radiated energy density from
unabsorbed and absorbed (but Compton-thin) AGNs using the hard X-ray
LF. The evolution of the LF is such that the luminosity density at $z>
5$ is similar to that at $z < 1$, but the lookback time is much
shorter, meaning the energy emitted at $z\geq 5$ is negligible. In
other words, the integral in Equation~\ref{eq:integral} has converged
by $z=5$. However, the LFs are not as well constrained in the range $3
\leq z \leq 5$, and the functional form chosen to represent the
evolution (e.g. a 2nd or 3rd order polynomial) can have unphysical
effects.  We therefore prefer to restrict the range over which the LFs
are integrated to $0 \leq z \leq 3$. The missing radiated energy is
expected to be $\lesssim$10\% (see Section~\ref{sec:lf}), and
therefore negligible compared to our uncertainties.

Due to the fact that the bolometric corrections are
luminosity-dependent and need to be included inside the integral when
calculating $U$, we quote here the product of $u$ and \cbol\, assuming
the mean value of \cbol. The uncertainty quoted is only due to $u$,
estimated by integrating different LFs, while the uncertainty in
\cbol\, is encoded in the prior P(\cbol).

For the B-band, due to the different survey areas and sensitivities,
we use the LFs of \citet {2000MNRAS.317.1014B} and \citet
{2004MNRAS.349.1397C} at $z< 1$ and of \citet{2003A&A...408..499W} in
the range $1 \leq z \leq 3$ (using both pure density and pure
luminosity evolution). The LF is integrated for all magnitudes
brighter than $M_{\rm B}<-18$ ($L_{\nu \rm B}$$\geq$8$\times$$10^{20}$
\whz). The estimated total energy from the B-band LFs is found to be
$u$\cbol$=$(1.1$\pm$0.2)$\times 10^{51}$~J~Mpc$^{-3}$, using the
bolometric corrections of \citet{2004MNRAS.351..169M}, and
$u$\cbol$=$(1.6$\pm$0.4)$\times 10^{51}$~J~Mpc$^{-3}$ when using
\cbol\, from \citet{2007ApJ...654..731H}.

If only the most powerful AGNs are considered (quasars, brighter than $M_{\rm
  B}$$=$-23 or $L_{\nu \rm B}$$=$7$\times$$10^{22}$ \whz), we find
$u$\cbol$=$(0.9$\pm$0.2)$\times10^{51}$ J~Mpc$^{-3}$ \citep[\cbol\,
from][]{2004MNRAS.351..169M} and $u$\cbol$=$(1.4$\pm$0.3)$\times
10^{51}$ J~Mpc$^{-3}$ \citep[\cbol\, from][]{2007ApJ...654..731H}.

We note all of the B-band LFs are likely to be missing the population
of reddened AGNs. These sources suffer from moderate amounts of
extinction in the visual band, $A_{\rm V}\sim1-5$, which suppresses
the AGN signatures at optical but not at near-infrared wavelengths
($\sim$1-2~$\mu$m, energies $\sim$0.5-1~eV).
\citet{2006ApJ...638...88B} have estimated these to comprise
$\sim$20\% of the unobscured population, so the above estimates of the
energy density include this uncertainty.

For the hard X-ray LF, we use the LFs of \citet{2003ApJ...598..886U}
and \citet{2008ApJ...679..118S}. The LF is then integrated over
$L_{\rm X}\geq$$3\times10^{34}$~W. The estimated radiated energy
density, which includes Compton-thin absorbed AGNs, is then
$u$\cbol$=$(4.5$\pm$0.1)$\times 10^{51}$~J~Mpc$^{-3}$ \citep[using
\cbol\, from][]{2004MNRAS.351..169M} and
$u$\cbol$=$(6.4$\pm$0.1)$\times 10^{51}$~J~Mpc$^{-3}$ \citep[\cbol\,
from][]{2007ApJ...654..731H}.

\subsubsection{{\rm P(\rtt)}}

As we have discussed in Section~\ref{sec:demo}, the value of \rtt\, inferred
amongst high-luminosity AGNs is $\sim$2, while amongst low-luminosity AGNs it
is $\sim$4. In Equations~\ref{eq:integral} and \ref{eq:densacc}, we have made
the simplification of keeping \rtt\, independent of luminosity and redshift
(and hence outside the integral).  The question of whether \rtt\, evolves or
not with $z$ is still unclear but it seems likely that it does vary with
luminosity (see Section~\ref{sec:demo}).

We will repeat our analysis using priors centred on \rtt$\sim$2 and 4, to show
the different values inferred for \eff. In addition we will also carry out the
analysis without including obscured AGNs, using
Equations~\ref{eq:integral_no_r} and \ref{eq:densacc_no_r}, which is
equivalent to using a prior with \rtt$=$0.

In the case of \rtt$\sim$2, inferred from the bright AGNs, we assign a
prior based on the estimate of \citet{2008ApJ...674..676M}.  This can be
approximated to a gaussian PDF with $\mu_{\langle R \rangle}=$1.7 and
$\sigma_{R}=$1.2 (their large uncertainty arising from small number
statistics).

Amongst low-luminosity AGNs, several studies suggest \rtt$\sim$4, but there is
no explicit estimate that we can use as a prior. For this case, we use a
$\delta$-function prior centred on 4. Due to the lack of scatter in \rtt, the
posterior PDF for \rtt$=$4 obtained from the $\delta$-function prior will be
artificially narrower than the posterior PDF assuming $\mu_{\langle R
  \rangle}=$1.7 and $\sigma_{R}=$1.2. This will be dealt with in
Section~\ref{sec:best}.

The hard X-ray LF includes unabsorbed and Compton-thin absorbed AGN, so that
only Compton-thick AGNs should be missing.  The luminosity-dependent absorbed
AGN fraction is also encoded in the LF.  We do not attempt to correct for the
missing Compton-thin population, so we are again using
Equation~\ref{eq:integral_no_r} and \ref{eq:densacc_no_r} instead of
~\ref{eq:integral} and \ref{eq:densacc}.

\subsubsection{{\rm P(\eff)}}

For the efficiency, we reflect our  ignorance by setting a prior in log
space, P(log$_{10}$[\eff]), which is flat in the range $-3.0
\leq$log$_{10}$[\eff]$\leq 0.0$.

\subsection{The local density of SMBHs}

For \densloc, we assume a value of $4.6^{+1.9}_{-1.4}$$\times 10^{5}$ \msol\,
Mpc$^{-3}$, following \citet{2004MNRAS.351..169M}.  The likelihood is then
assumed to be a gaussian distribution around this value
(Equation~\ref{eq:like}). We compare the total \densloc\, to the energy
density computed using all AGNs brighter than $M_{\rm B}<-18$ (for the B-band
LF) or $L_{\rm X}\geq$$3\times10^{34}$~W (hard X-ray LF).

For the case of \rtt$\sim$2, given that this value is motivated by luminous
obscured AGNs (quasars), we also consider only the bright end of the optical
LF (AGNs with $M_{\rm B}<-23$). We make the assumption that they are hosted
by the progenitors of present-day elliptical galaxies, so we compare their integrated
energy to the SMBH density in elliptical galaxies only.
\citet{2004MNRAS.351..169M} estimated the SMBH density in elliptical
galaxies to be 70\% of the total SMBH density, or $3.2^{+1.3}_{-1.0}$$\times
10^{5}$ \msol\, Mpc$^{-3}$.

\subsection{The posterior PDFs for \eff}\label{sec:poste}

Figure~\ref{fig:posteriors} (panels a, b, c and d) shows the resulting
posterior PDFs for log$_{10}$[\eff], given our parametrisation, given our
priors and given our assumed local density of SMBHs, \densloc. Panel (a) shows
the results from the B-band LFs without any correction for obscured AGNs
(\rtt$=$0), using our two sets of  values for \cbol. 

Panel (b) shows the result of using the hard X-ray LF. The value of $u$
derived from the hard X-ray LF (Section~\ref{sec:pu}) includes both X-ray
unabsorbed AGNs and X-ray absorbed Compton-thin AGNs, but not absorbed
Compton-thick AGNs.  

Panel (c) shows the results from the B-band LFs and the two different
bolometric corrections, with two different values of \rtt\, ($\sim$2 and 4)
used to correct for the obscured population (which includes the Compton-thick
AGNs as well as the Compton-thin ones).  Note that the posteriors for \eff\,
derived from \rtt$=$0 and \rtt$=$4 are artificially narrower than that for
\rtt$\sim$2, due to the $\delta$-function priors for \rtt. This is dealt with
in Section~\ref{sec:best}.

Panel d) shows the
posterior PDFs for the quasars. It is derived from the B-band LFs, integrating
only over AGNs brighter than $M_{\rm B}=-23$, considering only \rtt$\sim$2,
and comparing to the SMBH density in local elliptical galaxies.

\subsection{Combining  different estimates for \eff}\label{sec:best}

The estimates for \eff\, shown in Figure~2 have been derived using different
values for \cbol, and in the case of pannel~c, different values for \rtt. We
now wish to combine the estimates in each panel, to obtain our four estimates
for \eff\, for each of the three scenarios: a) considering only unobscured
quasars from the B-band LF b) considering the absorbed and unabsorbed AGNs
from the X-ray LF, but not the Compton-thick ones and c) attempting to
include all obscured and unobscured AGNs by using the B-band LF and correcting
by (1$+$\rtt) finally d) is the same as c) but only for the quasars.

In all four scenarios, we have used two different sets of bolometric
corrections. We have no strong reason to believe one set is more appropriate
than the other, but they do yield different values of \eff. 
Additionally, in scenario c, have derived estimates for \eff\, in the presence
of obscured AGNs based on two different priors for \rtt, one where
\rtt$\sim$2, the other with \rtt$=$4. Both of these values for \rtt\, are
believed to be accurate, but only appropriate at a certain luminosity.
Instead, we have used a fixed value of \rtt\, for the entire LF. We now wish
to improve our estimate, by considering how appropriate each of our 
estimates is, and combining them.

We follow the arguments of \citet{1997upa..conf...49P}, and consider our
different estimates of \eff\, from different combinations of the parameters
\cbol\, and \rtt\, as indepedent measurements. Although these inferred values
of \eff\, have been derived using the same method, and are therefore
apparently not independent, the reason they differ is due to the different
assumptions for \cbol\, and \rtt, which are themselves independent.  What we
are doing here is to estimate the most likely value of \eff, given the
independent parameters assumed.

 We must now assess the appropriateness of each estimate for \eff\, resulting
 from using each combination of parameters.  Each resulting estimate of \eff,
 \eff$_{i}$, will be considered a good estimate or a bad estimate, depending
 on whether the values of \cbol\, and \rtt\, assumed were appropriate or not.
 Bad estimates of \eff\, are simply those where the standard deviation has
 been underestimated.

 To each measurement \eff$_{i}$, we assign the probability $p_{i}$ that the
 choice of parameters was the most appropriate and probability (1-$p_{i}$)
 that it was not the most appropriate.  We here consider all $p_{i}$'s equal,
 simplifying to $p$.

The posterior probability for the efficiency \eff, given data $D$,  can then
be written as

\begin{eqnarray}\label{eq:press}
{\rm P({\rm log}_{10}[\langle \epsilon \rangle] | D)} = \nonumber \\{\rm P({\rm log}_{10}[\langle \epsilon \rangle]) \over
P(D)}\int  {\rm P({\it p})}\prod_{i}[p{\rm P}_{Gi} + (1-p){\rm P}_{Bi}]{\rm d}p
\end{eqnarray}

\noindent where P$_{Gi}$ and P$_{Bi}$ are the probability distribution
functions for good and bad measurements, derived from appropriate and
inappropriate parameter values \citep[see] [ for a derivation of this
result]{1997upa..conf...49P}.

We  marginalise over $p$. Thus, even if
we do not have any information on the probability of a term being correct or
not, and have no objective way of assigning values of $p$, we can still
proceed. We simply integrate over the entire range of values for $p$, weighted
by the prior P($p$).

To estimate ${\rm P({\rm log}_{10}[\langle \epsilon \rangle] | D)}$, the
posterior PDF for \eff\, given our different measurements, we assume a flat prior for
log$_{10}$[\eff].  ${\rm P({\rm log}_{10}[\langle \epsilon \rangle]) /
  P(D)}$ is in practice simply a normalisation term. For $p$, we assign a
prior flat in the range $0 \leq p \leq 1$, reflecting complete ignorance.

The good probability distribution functions P$_{Gi}$ are those where the
uncertainty is correctly estimated, so we use our individual posteriors for
each combination of parameters (the dashed, dotted or dash-dotted curves in
Figure~2 panel a, b, c or d). For the bad PDFs, P$_{Bi}$, we choose to assume they
would have been correct if they had quoted an uncertainty twice as large. We
rederive the curves, using exactly the same procedure and priors of
Section~\ref{sec:infer}, but with a final uncertainty which is made
artificially to be twice as large.

When combining the different estimates, we obtain an improved one that takes
into account not only the different mean parameter values, but also the
different uncertainties. For example, the potentially incorrect
(underestimated) width of the estimates for \eff\, derived from a
$\delta$-function prior at \rtt$=$4 has been dealt with.

The resulting combined posterior PDFs for \eff\, can be seen as the solid
curves in Figure~\ref{fig:posteriors} a, b, c and d. 
These are our best estimate for
the mean radiation efficiency, \eff, given each scenario.

\subsection{Results}\label{sec:results}

Our results for the different scenarios illustrate how the efficiency
estimated from the So{\l}tan argument varies as the population of obscured
AGNs is taken into account.  In all panels of Figure~2, the solid lines show
the result from combining the individual posterior PDFs assuming different
parameters. The uncertainties quoted below correspond to the gaussian
approximation of the solid lines in each figure. This gaussian approximation
is extremely accurate for the low-efficiency tail of the PDFs, but
underestimates the high-efficiency tail typically above the full-width
half-maximum.

When we consider only the unobscured AGNs, we find
log$_{10}$[\eff]$=$-1.77$^{+0.16}_{-0.11}$ (Figure~2a) or a relatively low
efficiency, \eff$=$1.7$^{+0.8}_{-0.4}$\%.  This is the efficiency derived when
no obscured AGNs are taken into account.

Using the hard X-ray LF
(Figure~2b), we find log$_{10}$[\eff]$=$-1.20$^{+0.15}_{-0.10}$ or
\eff$=$6.4$^{+2.6}_{-1.3}$\%. This estimate for \eff\, includes the absorbed
Compton-thin population, but is still missing the Compton-thick AGNs.

Figure~2c shows the result if we integrate over the optical LF and consider
the values of \rtt\, for high-luminosity and low-luminosity AGNs,
log$_{10}$[\eff]$=$-1.17$^{+0.11}_{-0.08}$ or \eff$=$6.7$^{+1.9}_{-1.1}$\%. We
consider this our overall best estimate for \eff, because it is expected to
include the Compton-thick AGNs and therefore to be the least incomplete in
terms of obscured AGNs.

Finally, when only obscured and unobscured quasars are considered we obtain
the solid curve in Figure~2d. Here we are using the optical LF integrated
above $M_{\rm B}=-23$ and assuming \rtt$\sim$2. We are also only comparing the
radiated energy density to the mass density of local SMBHs in elliptical
galaxies. The results can then be approximated as
log$_{10}$[\eff]$=$-1.27$^{+0.15}_{-0.11}$ or \eff$=$5.4$^{+2.2}_{-1.2}$\%.

The difference in efficiency between the B-band LF with \rtt$=$0 and the other
results shows the cosmological importance of obscured (or absorbed) AGNs.  In
Section~\ref{sec:spin}, we discuss how this affects the estimate of SMBH
spins, under a set of assumptions.  However, from the different results
obtained using different values of \cbol, visible in the different panels of
Figure~\ref{fig:posteriors}, we can also infer that the bolometric correction
is still a critical source of uncertainty. 

The value of \eff\, has been a matter of some debate, particularly whether its
value is above or below $\sim$10\%. In Section~\ref{sec:comp} we compare to
results obtained by other studies. We note that, from our best-estimate PDFs
from Figure~\ref{fig:posteriors}b, c and d, we find P(\eff$\leq$10\%) ranges
between 79 and 88\%. Hence, our results strongly favour values of \eff\, below
10\%.

\subsection{Comparison to previous estimates}\label{sec:comp}

Given the large uncertainties in \rtt, in \cbol\, and in \densloc, our
integrated approach is appropriate. Using this method, we obtain
\eff$\sim$7\%, and find our results consistent with those of \citet
{2004MNRAS.351..169M}, \citet{2004MNRAS.354.1020S} and
\citet{2008MNRAS.388.1011M} who used differential LFs (typically hard X-ray
LFs) and local SMBH mass functions, and applied extra corrections to account
for the Compton-thick population. 

Our results are partially in agreement with those of
\citet{2008arXiv0808.0759C}, who find \eff$\sim$8\% when considering
the AGNs leading to SMBHs with masses $<10^{8}$~M$_{\odot}$. However,
they also find \eff$\grtsim$18\% for $>10^{9}$~M$_{\odot}$, whereas we
have found that \eff$\sim$5\% when comparing quasars to the SMBH mass
density in elliptical galaxies. The SMBHs in ellipticals do not
necessarily map one-to-one with the SMBHs with $>10^{9}$~M$_{\odot}$
\citep[see e.g.][]{2004MNRAS.351..169M}, but our results suggest that
the mean radiation efficiency remains approximately constant.

We do not, however, find a good agreement with the results of
\citet{2008arXiv0808.3777Y}, who derive a value of \eff$\sim$16\%, or
those of \citet{2006ApJ...642L.111W} who used a completely independent
approach to estimate \eff$=$20-35\%.

Surprisingly, our result is also consistent with that of \citet
{2002MNRAS.335..965Y}, despite the fact that they did not include the
obscured AGN population. These authors used a lower value for
\densloc\, ($2.5\pm0.4$$\times 10^{5}$ \msol\, Mpc$^{-3}$) and also used
a higher value of \cbol, since they did not
remove the infrared component of the quasar SED. These factors
approximately cancel the effect of including the (1$+$\rtt) term,
explaining why our estimate of \eff\, is consistent with that of
\citet {2002MNRAS.335..965Y}.  

Using the intensity of the X-ray background, and assuming a typical redshift
of 2 for the sources dominating it, \citet {2002ApJ...565L..75E} estimated
\eff$\grtsim15$\%.  X-ray selected AGNs, however, have been observed to lie
predominantly at lower redshifts
\citep[e.g.][]{2003AJ....126..632B,2003ApJ...598..886U}. If a typical value of
$z=0.7-1$ is assumed instead, then the value of \eff\, derived following the
method of \citet {2002ApJ...565L..75E} is reduced to $\sim$8-10\%, in reasonable
agreement with our best estimate.

\section{Estimating the mean SMBH spin}\label{sec:spin}

The rotation of a black hole determines the radius of the innermost
stable circular orbit (ISCO) and therefore its binding energy. The
rotation is measured in terms of the parameter $a\equiv{J\over c
  m_{\bullet}}$ (in units of ${G m_{\bullet} \over c^{2}}$), where $J$
is the angular momentum, $m_{\bullet}$ is the mass of the SMBH, $G$ is
the gravitational constant and $c$ is the speed of light.  The
accretion efficiency ($\eta$, equal to 1 minus the binding energy and
shown in Figure~\ref{fig:spin}) is therefore determined by $a$.

The standard model for a thin accretion disk assumes no stresses are
present between the ISCO and the Schwartzschild radius and neglects
the effects of radiation from matter falling through this section, as
well as assuming that any heat dissipated in the disk is immediately
radiated away \citep{1973blho.conf..343N}. These assumptions then lead
to $\epsilon = \eta(a)$, so that the mean radiation efficiency can be
used to estimate the mean SMBH spin.

Relaxing these assumptions has two important consequences, $\epsilon$
is altered and, through accretion alone the SMBH might reach an
equilibrium spin value lower than $a={G m_{\bullet} \over c^{2}}$.

When radiation inside the ISCO is considered, the black hole captures
preferentially photons with opposite angular momentum, which limits
the final spin to $a=$0.998${G m_{\bullet} \over c^{2}}$
\citep{1970PhRvD...1.2721G,1974ApJ...191..507T}. Magnetic fields in
the region inside the ISCO can transport angular momentum outwards
\citep{1977MNRAS.179..433B}, possibly further limiting the equilibrium value
of $a$ \citep[][ estimate a limit of $\sim$0.9${G m_{\bullet} \over
  c^{2}}$]{2004ApJ...602..312G,2005ApJ...622.1008K}.

The assumption that $\epsilon = \eta$ requires that no torques are present
within the ISCO, and that the matter ``plunges'' into the black hole too
quickly for any radiation to be significant. In this case, no energy can be
extracted once matter has crossed the ISCO. However, the infalling matter can
exhert a torque through magnetic fields, so that work is still done on the
disk even after crossing the ISCO. The dissipation of this work can then lead
to a higher radiative efficiency
\citep[e.g.][]{2008MNRAS.tmp.1003B,2008arXiv0808.3140N}.

In the following section, we discuss our estimates of $\langle a
\rangle$ from \eff\, assuming $\epsilon = \eta$, but from the
discussion above these values should be considered as upper limits,
since $\epsilon$ is really expected to be $>\eta(a)$. For the case of
SMBH and disk co-rotating with $a\sim0.9 {G m_{\bullet} \over c^{2}}$,
the ISCO is very close to the Schwartzschild radius, so that there is
little time to radiate in the region inside of the ISCO. In addition,
the gravitational redshift diminishes the energy of the emergent
radiation. For this case, \citet{2008arXiv0808.3140N} have found that
$\epsilon\sim 1.06 \eta$, so the difference is negligible compared to
our uncertainties. However, for lower values of $a$ the difference
between $\epsilon$ and $\eta$ might be larger, even of order unity
\citep{2008MNRAS.tmp.1003B}.

Our value of \eff\, derived in the absence of obscured AGNs
(\eff$=$1.7$^{+0.8}_{-0.4}$\% for \rtt$=$0) is consistent with SMBHs having no
rotation during periods of accretion (\eff$<$5.7\% so that $a=0$). An
important point is therefore that when obscured AGNs are not included, the
derived efficiency suggests that SMBHs were not spinning during periods of
accretion.

When we do include the obscured AGNs, we find that values of \eff$>$5.7\% are
likely, so that the possibility that the SMBHs are rotating must be
considered.

Maximally rotating black holes have $a={G m_{\bullet} \over c^{2}}$
with $\eta=$42\% \citep{1970Natur.226...64B}, although in the standard
accretion disk model they are only expected to reach $a=0.998{G
  m_{\bullet} \over c^{2}}$ at most, corresponding to $\eta=$30\%
\citep{1974ApJ...191..507T}.  From our posterior PDFs,
P(\eff$\geq$30\%) ranges from 0.3\%\footnote{These probabilities have
  been calculated from the actual PDFs, not the gaussian approximation
  which underestimates the probability in the tails of the PDFs.} to
2\%$^{2}$.  Thus, under our parametrisation, we can rule out that
SMBHs are, on average, maximally rotating with $\geq$98\%
confidence. More recent work suggests the upper limit on $a$ might be
lower, $\sim$0.9${G m_{\bullet} \over c^{2}}$, and with a radiative
efficiency $\sim$15\% \citep{2008arXiv0808.3140N}. Again, from our
posterior PDFs, P(\eff$\geq$15\%) ranges from 2\% to 8\%, so we rule
out $\langle a \rangle \geq 0.9 {G m_{\bullet} \over c^{2}}$ at the
$\geq$92\% confidence.

For spinning black holes, $\langle \eta \rangle$ will vary depending on whether the ISCO is
co-rotating or counter-rotating with respect to the black hole spin (see
Figure~\ref{fig:spin}). To estimate the mean value of $a$  two possible limiting scenarios can be envisaged:

{\it A) The SMBHs grow from the seed in one long episode of accretion}. In
this case one single accretion disk is expected to be able to provide enough
mass to turn the seed into a SMBH. The
angular momentum of the seed is negligible compared to that of the accretion
disk, so the disk will inevitably force the black hole to co-align
\citep{1996MNRAS.282..291S,2005MNRAS.363...49K}. It is therefore safe to
estimate the efficiency assuming the black hole and ISCO are
always co-rotating. In this case, our best estimate of \eff$=$6.7$^{+1.9}_{-1.1}$\%
corresponds to SMBHs rotating with $\langle a \rangle=0.25^{+0.30}_{-0.25}{G m_{\bullet} \over
  c^{2}}$ during periods of accretion (Figure~\ref{fig:spin}, solid line).

{\it B) The SMBHs grow in many shorter periods of accretion}. The implication
here is that the angular momenta of the accretion disk and black hole are
comparable and are also initially randomly oriented with respect to each
other. Depending on the initial configuration and ratio of angular momenta,
the black hole and disk will either co-align or counter-align
\citep{2005MNRAS.363...49K,2006MNRAS.368.1196L}. It is expected that co- and
counter-aligned accretion is equally probable, so when estimating $a$, a mean $\langle \eta(a) \rangle$ from co- and counter-rotating
orbits should be used \citep[the symmetrised curve suggested by][shown in
Figure~\ref{fig:spin} by the dashed line]{2008MNRAS.385.1621K}. In this case,
for our best estimate of \eff\, we inferr 
$\langle a \rangle=0.59^{+0.25}_{-0.59}{G m_{\bullet} \over c^{2}}$.

Mergers involving SMBHs are expected to prevent $a$ from reaching the
highest values \citep[see the work of ][ who find half of their
parameter space will lead to $a \leq 0.4{G m_{\bullet} \over
  c^{2}}$]{2003ApJ...585L.101H}. For both accretion scenarios, and for
$a\lesssim0.8{G m_{\bullet} \over c^{2}}$, the dependence of
$\eta$ on $a$ is weak so that it is difficult to constrain
$\langle a \rangle$ from our estimate of \eff.  However, we have found
that maximally rotating SMBHs are not favoured since
P(\eff$\geq$30\%)$\leq$2\%, and our best estimates for $a$ are in
reasonable agreement with the results of \citet{2003ApJ...585L.101H}.

Our estimates for $\langle a \rangle$ are nevertheless also consistent with
the value of $a\geq0.5{G m_{\bullet} \over c^{2}}$ estimated for the SMBH in
the Milky Way, Sagittarius A$^{\star}$ \citep[][ who explained the short
near-infrared flare period of 17 minutes as being close the ISCO for
co-rotating disk and SMBH]{2003Natur.425..934G}. A SMBH with $a\geq0.5{G
  m_{\bullet} \over c^{2}}$ and a prograde orbit has an accretion efficiency
$\eta\geq$8\%. Thus, the limit on $a$ inferred for the individual SMBH
Sagittarius A$^{\star}$ is in good agreement with our best estimate of the
mean radiation efficiency, \eff, assuming the standard thin disk model and
ignoring the magnetic fields.

Indeed, we remind the reader that we have estimated $\langle a \rangle$
assuming $\epsilon=\eta$, but at the beginning of this Section we have argued
for $\epsilon>\eta$. Hence, our estimates of $\langle a \rangle$ can be
considered upper limits, since for a given true value of $a$,
$\epsilon>\eta(a)$
\citep{2005ApJ...622.1008K,2008arXiv0808.3140N,2008MNRAS.tmp.1003B}. For
$a\sim0.9 {G m_{\bullet} \over
    c^{2}}$, it seems that $\epsilon$ is only slightly larger than $\eta$, but
for lower values of $a$ the difference is expected to be larger, since the
time available for radiation is longer and the gravitational redshift is less
severe.  This only strengthens our main conclusion that our inferred value of
\eff\, does not require accreting SMBHs to be rotating particularly rapidly,
on average, during periods of accretion.

\section{The mean $e$-folding time}\label{sec:efold}

Given our assumption that AGNs grow by accretion, their mass will grow
exponentially provided that the luminosity is proportional to the mass. Thus,
if AGNs radiate at a fraction $\lambda$ of their Eddington luminosity, then
they will grow exponentially with an $e$-folding time:

\begin{equation}
\tau = {\epsilon \over \lambda (1-\epsilon)}{c \sigma_{\rm T} \over  4
    \pi G m_{\rm p} }
\end{equation}

\noindent where $m_{\rm p}$ is the mass of the proton and $\sigma_{T}$ is the
  Thomson electron scattering cross-section. Our estimate of \eff\, is
  completely independent of $\lambda$, and it is only to estimate $\tau$ that
  we require an estimate of the Eddington ratio $\lambda$.

  Estimates of $\lambda$ exist for unobscured quasars, derived from their
  bolometric luminosities  and virial-mass estimates for the SMBH
  mass.  \citet{2004MNRAS.352.1390M} found $\lambda\sim$0.15-0.5 in the range
  $0.1\leq z \leq 2.1$, with significant scatter.  We approximate the 
 distribution of
  ${\rm log}_{10}[\lambda]$ to a gaussian with $\mu_{{\rm
      log}_{10}[\lambda]}$$=$-0.6 and $\sigma_{{\rm
      log}_{10}(\lambda)}$$=$0.4.

  When combined with our estimate of \eff\, for quasars (5.4$^{+2.2}_{-1.2}$\%),
  this leads to an estimate of log$_{10}$[$\langle \tau
  \rangle$/Myr]$=2.0\pm0.4$, or $\langle \tau \rangle=100^{+151}_{-60}$~Myr.
  This value is consistent with (although at the high end of) other current
  constraints from quasar clustering measurements, from arguments based on
  coevolution of quasars, SMBHs and host galaxies, from arguments of
  simultaneous triggering of quasars and starbursts and from the lengths of
  radio jets. These estimates all suggest $\tau$ is in the range 10-100~Myr
  \citep[see][ for a review]{2004cbhg.symp..169M}.

  Such a long value of $\tau$ suggests SMBHs require a long time to build up
  their mass. At $z=6.4$, SMBH masses of 2-6$\times10^{9}$~\msol\, have been
  estimated \citep{2003ApJ...587L..15W,2003ApJ...594L..95B}, when the Universe
  was only 840~Myrs old. This suggests that the seeds that formed the $z=6$
  quasars might have been as large as $\sim10^{5}-10^{6}$~\msol.  

  The estimates of the SMBH mass and particularly our estimate of the
  $e$-folding time have very large uncertainties and the quasars
  at $z=6.4$ could have  shorter $e$-folding times. They might
  be accreting closer to the Eddington limit than $\lambda$$\sim$0.25, the
  mean of the $0.1 \leq z \leq 2.1$ quasars \citep[][]{2004MNRAS.352.1390M},
  or radiating less efficiently than $\sim$7\% . This would then imply
  significantly smaller seeds. For example, if $\tau\sim$60~Myr (40~Myr), the
  seeds could be $\sim$$10^{3}$~\msol\, ($\sim$1~\msol).

\section {Summary}\label{sec:conclusion} 

We have revisited the mean radiation efficiency from accretion onto SMBHs,
\eff, quantifying the effect introduced by taking into account a large
population of obscured AGNs.  We estimate the cumulative energy density of the
AGN population from the unobscured AGN LF by applying appropriate bolometric
corrections and correcting for the obscured population. The efficiency \eff\,
is estimated by comparing this energy density to the SMBH mass density in the
local Universe.

When no obscured AGNs are included, we find \eff$=$1.7$^{+0.8}_{-0.4}$\%  which
does not require SMBHs to rotate at all during periods of accretion ($\langle a\rangle$$=0$).

Using the hard X-ray LF, which includes absorbed Compton-thin AGNs, we find
\eff$=$6.4$^{+2.6}_{-1.3}$\%. Using the B-band LFs, correcting for the
obscured population (which includes the Compton-thick AGNs), we find
\eff$=$6.7$^{+1.9}_{-1.1}$\%, which we consider to be our best estimate.

We have also derived \eff\, for the powerful AGNs only (the quasars), and find  \eff$=$5.4$^{+2.2}_{-1.2}$\%.

Depending on whether the SMBHs and accretion disks are assumed to be always
co-rotating, or instead to counter-rotate in half of the cases, we estimate
the rotation parameter during periods of accretion to be $\langle a
\rangle=0.25^{+0.30}_{-0.25}{G m_{\bullet} \over c^{2}}$ or $\langle a
\rangle=0.59^{+0.25}_{-0.59}{G m_{\bullet} \over c^{2}}$, respectively

Efficiencies $\geq$30\% ($\geq$15\%) are ruled out at $\geq$98\% ($\geq$92\%)
confidence, so the SMBHs are not, on average, maximally rotating during the
periods of accretion.  Efficiencies $\leq$10\% are favoured, with 79\%-88\% of
the probability density in this region.

Finally, combining our best estimate for \eff\, with estimates of the
Eddington ratio for unobscured quasars yields an estimate of the $e$-folding
time of $\langle \tau \rangle=100^{+151}_{-60}$~Myr. This 
suggests that the seeds of $z=6$ quasars might have been very large,
 potentially as large as $\sim10^{6}$~\msol.  However, the large
uncertainties in $\tau$ translate into order-of-magnitude uncertainties for
the masses of the seeds, which could be as small as $\sim1$~\msol.

 \acknowledgements 

 We are grateful to John Miller, Andrew King and Giuseppe Lodato for useful
 discussions, to Ross McLure for access to the distribution of Eddington
 ratios, to Fabio Fontanot and Mark Sargent for comments on the manuscript and
 to the anonymous referee for suggestions that improved the article
 significantly.

\bibliographystyle{apj}

\clearpage 

\begin{figure} 
\plottwo{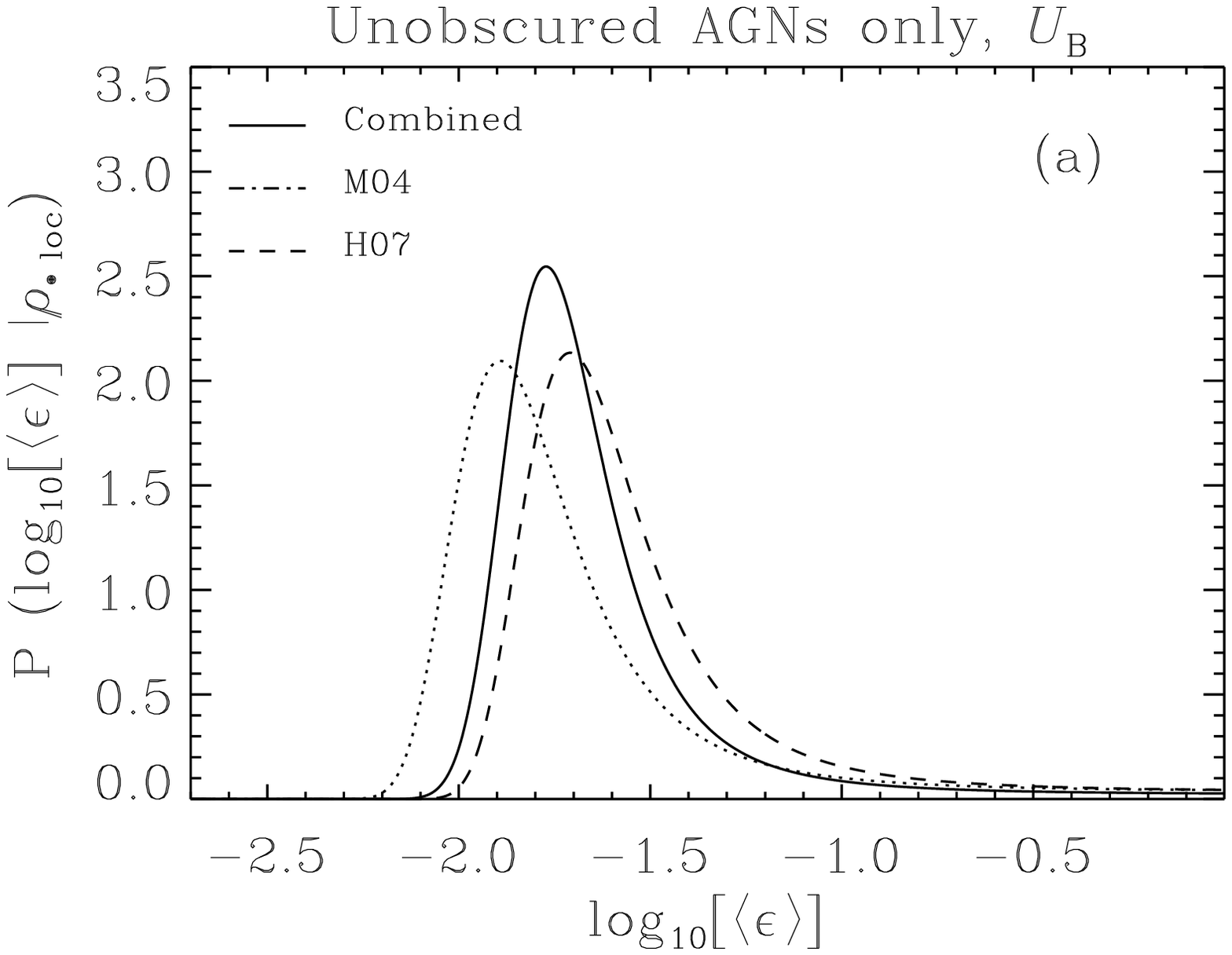}{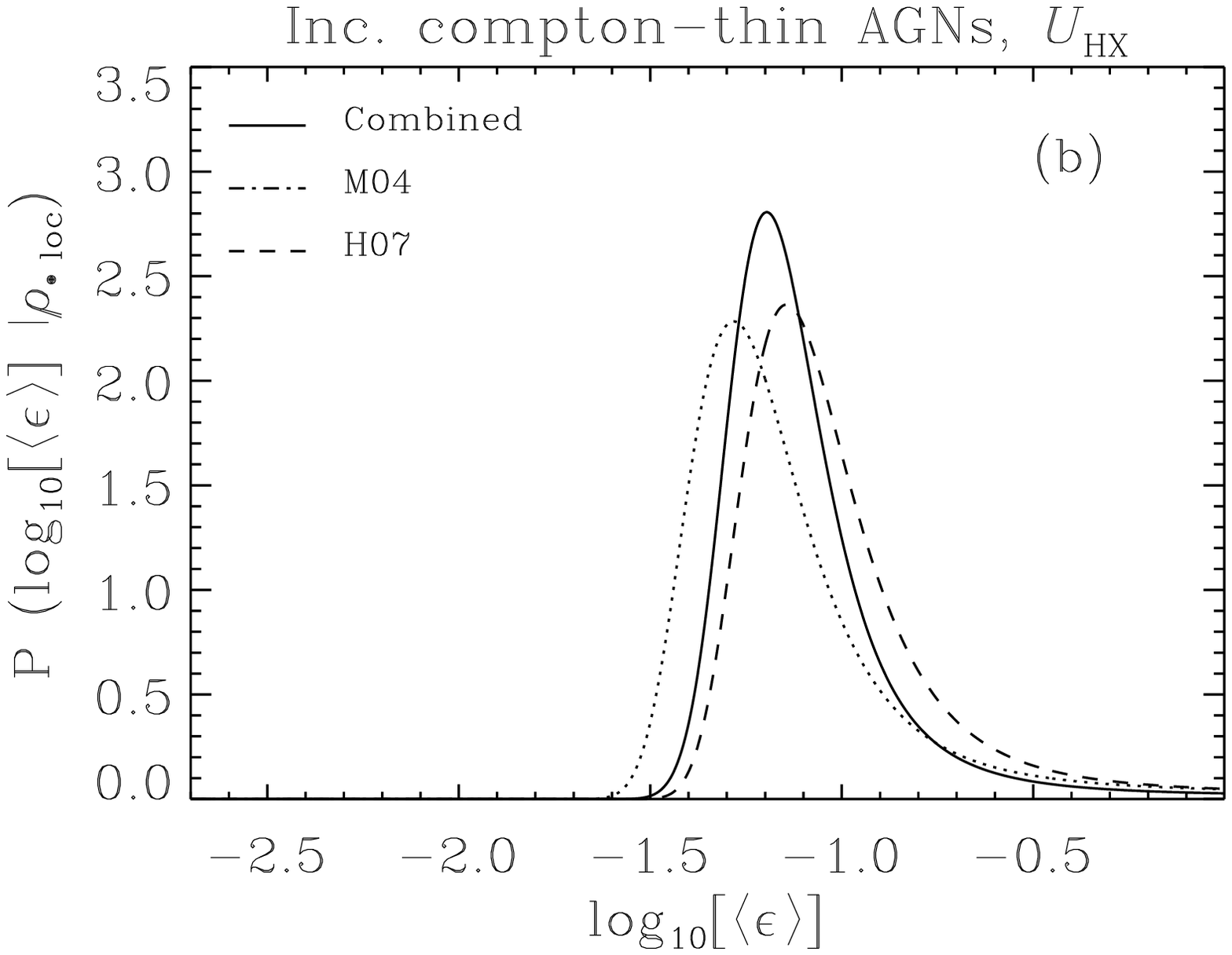}
\caption{\noindent Posterior PDF for \eff\, derived from the optical
  LFs for unobscured AGNs only (a) and using the hard X-ray LF, with
  no correction for Compton-thick absorbed AGNs (b). The posteriors
  derived from the bolometric corrections of
  \citet{2004MNRAS.351..169M} are marked as 'M04', whereas those from
  \citet{2007ApJ...654..731H} are marked as 'H07'. The combined
  posteriors, derived in Section~\ref{sec:best} are shown as the solid
  curve. For panel (a), the optical LF is integrated over all
  magnitudes brighter than $M_{\rm B}=-18$ ($L_{\nu \rm
    B}$$\geq$8$\times$$10^{20}$ \whz), whereas for panel (b) the LF
  has been integrated over all $L_{\rm X}\geq3\times10^{34}$~W.}
\label{fig:posteriors} 
\end{figure} 

\addtocounter{figure}{-1}
\begin{figure} 
\plottwo{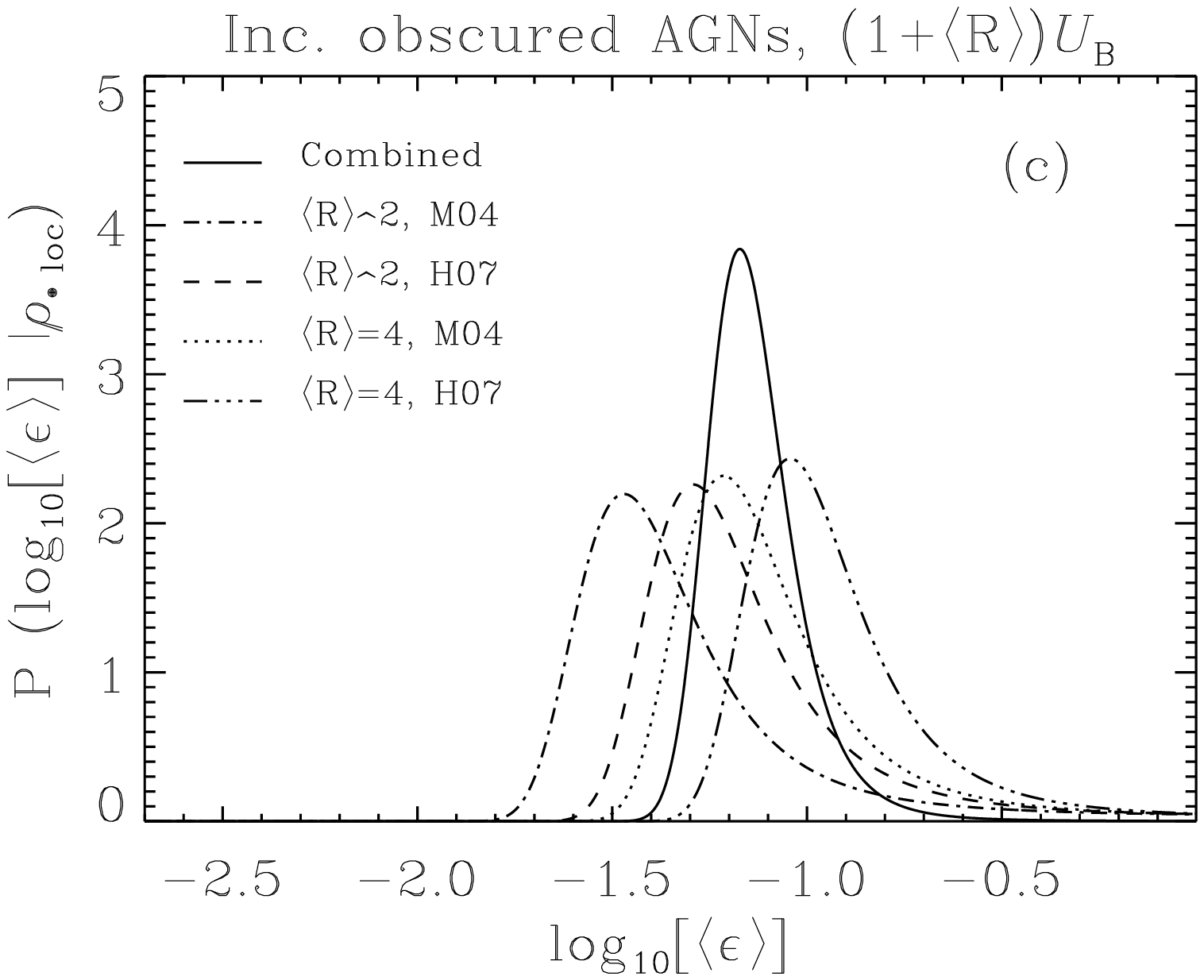}{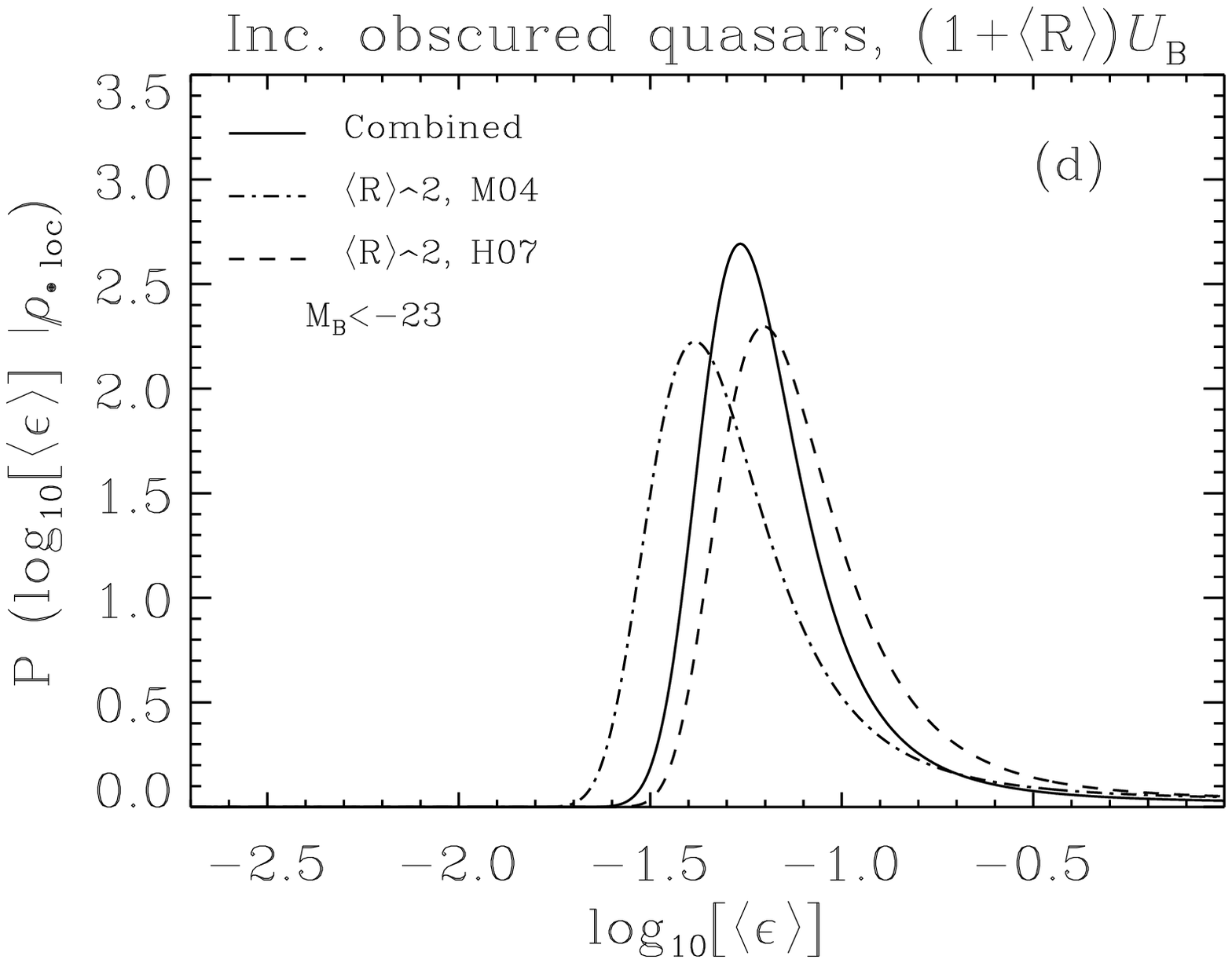}
\caption{ Continued. Posterior PDFs for \eff\, derived from the optical LFs
  correcting for the obscured AGNs using \rtt. In panel (c), the optical LF is
  integrated over all magnitudes brighter than $M_{\rm B}=-18$, and two
  different values of \rtt\, are used (\rtt$\sim$2 and $=$4). We consider this
  our best estimate for \eff. In panel (d), the LF has been
  integrated over all magnitudes brighter than $M_{\rm B}=-23$, and only
  \rtt$\sim$2 is used, to estimate the value of \eff\, from the bright AGN
  population (the quasars). }
\end{figure}

\begin{figure} 
\plotone{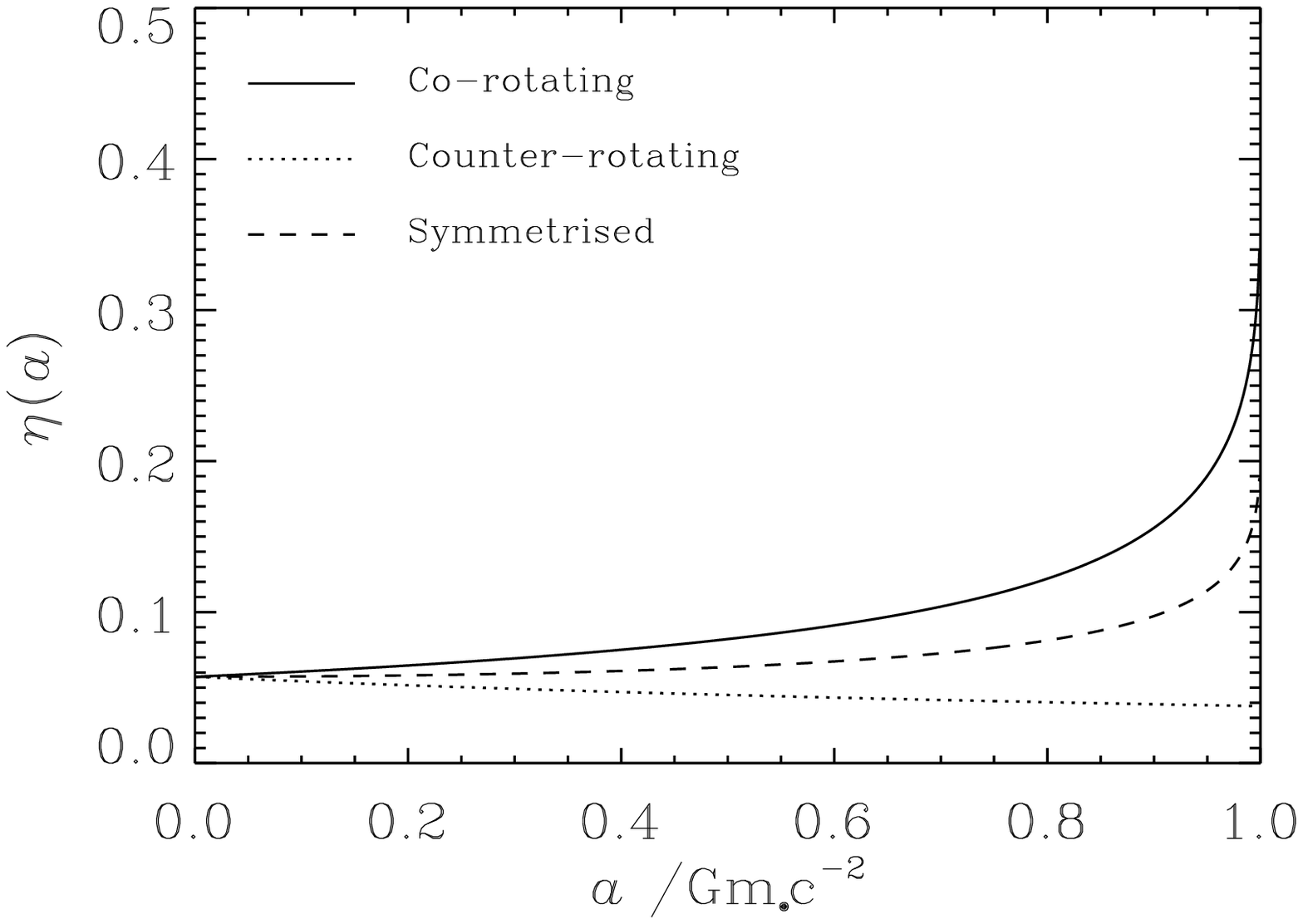}
\caption{\noindent Accretion efficiency $\eta$ as a function of rotation
  parameter $a$. The solid curve represents the case of the SMBH co-rotating
  with the inner part of the accretion disk, so the innermost stable circular
  stable orbit (ISCO) is prograde. The dotted line represents the case of
  counter-rotating SMBH and inner accretion disk (the ISCO is retrograde). The dashed curve represents the mean of both curves
  \citep [the ``symmetrised'' curve, see ][]{2008MNRAS.385.1621K}. 
}
\label{fig:spin} 
\end{figure}

\label{lastpage} 
 
\end{document}